\documentclass[twocolumn,amstext,amsmath,amssymb,prc,superscriptaddress,floatfix,showpacs,nofootinbib]{revtex4}
\usepackage[dvips]{epsfig}
\usepackage{slashed}
\usepackage{mathrsfs}
\usepackage{graphicx}
\usepackage{xcolor}
\usepackage{hyperref}
\usepackage{dcolumn}
\usepackage{subfigure}
\usepackage{xspace}
\usepackage{bbm}

\definecolor{heidelbeer}{rgb}{0.5,0,0.5}

\newcommand{\tr}[1]{\operatorname{Tr}\left[#1\right]}
\newcommand{\re}[1]{\operatorname{Re}\left[#1\right]}
\newcommand{\im}[1]{\operatorname{Im}\left[#1\right]}

\newcommand{\statbracket}[1]{\big\langle#1\big\rangle}

\graphicspath{
{./}
{./Figures/}
}
\begin{document}
\title{Simulating fermion production in $1+1$ dimensional QED}

\author{F. Hebenstreit}
\email[]{f.hebenstreit@thphys.uni-heidelberg.de}
\affiliation{Institut f\"{u}r Theoretische Physik, Universit\"{a}t Heidelberg,
  Philosophenweg 16, 69120 Heidelberg, Germany}

\author{J. Berges}
\email[]{j.berges@thphys.uni-heidelberg.de}
\affiliation{Institut f\"{u}r Theoretische Physik, Universit\"{a}t Heidelberg,
  Philosophenweg 16, 69120 Heidelberg, Germany}
\affiliation{ExtreMe Matter Institute EMMI, GSI Helmholtzzentrum,
  Planckstra\ss e 1, 64291 Darmstadt, Germany}

\author{D. Gelfand}
\email[]{d.gelfand@thphys.uni-heidelberg.de}
\affiliation{Institut f\"{u}r Theoretische Physik, Universit\"{a}t Heidelberg,
  Philosophenweg 16, 69120 Heidelberg, Germany}

\begin{abstract}
We investigate fermion--anti-fermion production in $1+1$ dimensional QED using real-time lattice techniques. 
In this non-perturbative approach the full quantum dynamics of fermions is included while the gauge field dynamics can be accurately represented by classical-statistical simulations for relevant field strengths. We compute the non-equilibrium time evolution of gauge invariant correlation functions implementing 'low-cost' Wilson fermions. 
Introducing a lattice generalization of the Dirac-Heisenberg-Wigner function, we recover the Schwinger formula in $1+1$ dimensions in the limit of a static background field. We discuss the decay of the field due to the backreaction of the created fermion--anti-fermion pairs and apply the approach to strongly inhomogeneous gauge fields. The latter allows us to discuss the striking phenomenon of a linear rising potential building up between produced fermion bunches after the initial electric pulse ceased.  
\end{abstract}
\pacs{11.10.Kk, 11.15.Ha, 12.20.Ds}
\maketitle


\section{Introduction}

It has already been pointed out in the early days of quantum physics that the vacuum of quantum electrodynamics (QED) becomes unstable against the formation of many-body states in the presence of strong external electromagnetic fields, manifesting itself as the creation of electron-positron pairs by the Schwinger mechanism \cite{Sauter:1931zz,Heisenberg:1935qt,Schwinger:1951nm}.
Nevertheless, this fundamental quantum effect has not been experimentally observed so far as it has not been possible to generate the required electromagnetic field strengths in a laboratory.
However, due to the rapid development of laser technology during the last decades an experimental veriﬁcation of electron-positron pair production in the focus of high-intensity laser pulses comes into reach. 

Vacuum pair production in an applied uniform electric field of strength $E_0$ may be viewed as a quantum process in which virtual electron-positron dipoles can be separated to become real pairs once they gain the binding energy of $2mc^2$.
However, there will be strong spatial and temporal inhomogeneities of the electromagnetic field in realistic situations as envisaged in upcoming high-intensity laser experiments. 
The theoretical description of such a non-perturbative phenomenon in quantum field theory out of equilibrium is a demanding task and very little is known so far for realistic scenarios. 
Most current approaches assume the electromagnetic field as being an external one with a one-dimensional inhomogeneity, so that the problem of particle production can be mapped onto a one-dimensional quantum mechanical scattering problem \cite{Brezin:1970xf,Popov:1972}. This approach neglects, in particular, the backreaction of the created fermion--anti-fermion pairs on the electromagnetic field.
This is closely related to kinetic descriptions in terms of a momentum dependent distribution function of pairs in collisionless (Vlasov) approximations \cite{Kluger:1992gb,Schmidt:1998vi,Alkofer:2001ik,Blaschke:2005hs,Tanji:2008ku,Hebenstreit:2009km}. 
For multi-dimensional inhomogeneities, more advanced approaches such as semi-classical approximations of the vacuum effective action \cite{Dunne:2006ur} or the Dirac-Heisenberg-Wigner phase space formulation \cite{BialynickiBirula:1991tx,Hebenstreit:2010vz} have been applied. However, to describe strongly inhomogeneous field configurations including the full backreaction of the produced particles remains a theoretical challenge. In view of the potential experimental applications it is crucial to devise new theoretical methods which can deal with this situation. 

In this work we propose to use real-time lattice gauge theory techniques \cite{Aarts:1998td} to compute fermion--anti-fermion pair production in QED. In this non-perturbative approach the full quantum dynamics of fermions is included while the gauge field dynamics can be accurately represented by classical-statistical simulations for relevant field strengths. As the inclusion of dynamical fermions can become numerically very expensive, the real-time evolution of fermions is taken into account by means of a low-cost fermion algorithm \cite{Borsanyi:2008eu}. 

As an example and in order to compare with established continuum results, we apply these techniques to QED in $1+1$ dimensions -- the massive Schwinger model \cite{Coleman:1975pw,Coleman:1976uz}.
Introducing a lattice generalization of the Dirac-Heisenberg-Wigner function, we show that the simulations accurately reproduce the results described by the Schwinger formula in the limit of a static background field. 
We discuss the decay of the field due to the backreaction of the created fermion--anti-fermion pairs and apply the approach to strongly inhomogeneous gauge fields. 
For these fields we compute for the first time the backreaction of the created pairs on the gauge fields. 
Most strikingly, we find that a self-consistent electric field between the produced fermion and the anti-fermion bunch builds up for times exceeding the initial pulse duration. 
The two bunches consisting of particles and anti-particles act as a capacitor, creating a homogeneous electric field between them, which can be represented in terms of a linear rising potential.

From the point of view of theoretical developments, it is important to note that very similar questions are addressed in physics of the early universe or in collision experiments of heavy nuclei. Non-equilibrium particle creation from large coherent fields has been extensively studied in the context of inflaton dynamics \cite{Kofman:1994rk,Prokopec:1996rr,Berges:2002cz} and non-Abelian gauge field theory \cite{Fujii:2009kb,Berges:2011sb}.  
The production of fermion--anti-fermion pairs has been mainly investigated based on semi-classical descriptions using the Dirac equation coupled to time-dependent background fields \cite{Baacke:1998di,Greene:1998nh,GarciaBellido:2000dc,Gelis:2004jp}. Going beyond these approximations, studies based on two-particle irreducible (2PI) effective action techniques \cite{Berges:2009bx,Berges:2010zv} showed that quantum effects can dramatically affect  the far-from-equilibrium production of fermion pairs. Recently, pair production from scalar inflaton decay \cite{Berges:2010zv} as well as baryogenesis \cite{Saffin:2011kc,Saffin:2011kn} has been studied using similar lattice field theory methods as employed in this work. 
In particular, the results of Refs.~\cite{Berges:2010zv,Saffin:2011kn} provide a proof of principle that real-time lattice simulations with Dirac fermions are indeed feasible in $3+1$ dimensions. They also have been tested \cite{Berges:2010zv} against calculations based on real-time 2PI effective action techniques \cite{Berges:2004yj,Berges:2002wr} in their range of applicability.  

This paper is organized in the following way:
In Sec.~\ref{sec:theory} we briefly review the low-cost fermion algorithm and derive the real-time lattice equations of motion for the massive Schwinger model.
Additionally, we construct a lattice generalization of the Dirac-Heisenberg-Wigner function which is subsequently used as a read-out tool for fermionic distributions. 
In Sec.~\ref{sec:results} we first apply this formalism to a static electric background field and compare to the Schwinger formula. 
We then discuss the decay of the background field due to the backreaction of the created fermion--anti-fermion pairs.
As a second example, we investigate the pair creation process in the presence of a space- and time-dependent electric field.
In Sec.~\ref{sec:conclusion} we conclude and give an outlook. 


\section{Real-time lattice gauge theory}
\label{sec:theory}

\subsection{Continuum formulation}
\label{sec:theory_low}

We consider QED in $1+1$ dimensions, which is defined in the continuum by the action
\begin{equation}
 \label{fmla_th_action_cont}
 \mathcal{S}=\int{d^2x\left(\bar{\psi}[i\gamma^\mu D_\mu-m]\psi-\frac{1}{4}\mathcal{F}^{\mu\nu}\mathcal{F}_{\mu\nu}\right)} \ , 
\end{equation}
with the covariant derivative $D_\mu=\partial_\mu+ie\mathcal{A}_\mu$ ensuring gauge invariance of the action under local $U(1)$ transformations
\begin{equation}
 \label{fmla_th_gauge_cont}
 \psi\to\psi e^{ie\Lambda} \quad , \quad \mathcal{A}_\mu\to \mathcal{A}_\mu-\partial_\mu\Lambda \ .
\end{equation}
Here $\mu=0,1$ as space-time is only two-dimensional with $x^0\equiv t$ and $x^1\equiv x$. The field strength tensor $\mathcal{F}^{\mu\nu}=\partial^\mu \mathcal{A}^\nu-\partial^\nu \mathcal{A}^\mu$ possesses only one non-trivial component which is regarded as the electric field:
\begin{equation}
 \label{fmla_th_el_cont}
 \mathcal{F}^{10}=-\mathcal{F}^{01}=E(x,t) \ .
\end{equation}
We will frequently consider temporal axial gauge with $\mathcal{A}_0(x,t) = 0$ and simply denote the spatial component of the vector potential as $\mathcal{A}(x,t)$.\footnote{We note that this incomplete gauge choice leaves a residual gauge invariance under time-independent gauge transformations.} 
One observes that the electric field $E(x,t)$ is the canonical momentum conjugate to $\mathcal{A}(x,t)$. 

The Dirac algebra is composed of two Dirac gamma matrices only:
\begin{equation}
 \{\gamma^\mu,\gamma^\nu\}=2g^{\mu\nu}	\qquad \mathrm{with} \qquad (\gamma^\mu)^\dagger=\gamma^0\gamma^\mu\gamma^0 \ ,
\end{equation}
with $g^{\mu\nu}=\operatorname{diag}(1,-1)$. 
This algebra may be represented in terms of the first two Pauli matrices $\gamma^0\equiv\sigma_1$ and $\gamma^1\equiv-i\sigma_2$.
Moreover, the chirality matrix
\begin{equation}
 \{\gamma^\mu,\gamma_5\}=0 \qquad \mathrm{with} \qquad (\gamma_5)^\dagger = \gamma_5 \ , \ (\gamma_5)^2=1 \ ,
\end{equation}
can be defined in terms of the third Pauli matrix $\gamma_5\equiv\sigma_3$.
As a consequence, the spinors $\psi$ and $\bar{\psi}$ are two-component field operators, obeying the equal-time anticommutation relation:
\begin{equation}
 \label{fmla_th_anticomm}
 \{\psi(x,t),\bar{\psi}(y,t)\}=\gamma^0\delta(x-y) \ .
\end{equation}

\subsubsection{Time evolution equations}

In general, in the classical-statistical theory observables are calculated as ensemble averages of solutions of Maxwell's equation 
\begin{equation}
\label{fmla_th_eomfs}
\partial_\mu \mathcal{F}^{\mu\nu}(x,t)= \langle j^\nu(x,t) \rangle
\end{equation}
starting from different canonical field variables at initial time $t_0$, here $\mathcal{A}_{t_0}(x) = \mathcal{A}(x,t_0)$ and $E_{t_0}(x)=E(x,t_0)$.
The values for the canonical field variables at initial time are distributed according to a normalized phase-space density functional $W[\mathcal{A}_{t_0},E_{t_0}]$, such that an observable $\langle O \rangle$ is given by \cite{QvsCS1,QvsCS2}:
\begin{equation}
\label{eq:expCS}
\langle O\rangle = \int D\mathcal{A}_{t_0} DE_{t_0}\,W[\mathcal{A}_{t_0},E_{t_0}]\,O_{\text{cl}}[\mathcal{A}_{t_0},E_{t_0}]\;. 
\end{equation} 
Here $O_{\text{cl}}[\mathcal{A}_{t_0},E_{t_0}] = \int D\mathcal{A}\,O[\mathcal{A}]\,\delta(\mathcal{A}-\mathcal{A}_{\text{cl}}[\mathcal{A}_{t_0},E_{t_0}])$,
where $\mathcal{A}_{\text{cl}}[\mathcal{A}_{t_0},E_{t_0}]$ is the solution of the classical field equation (\ref{fmla_th_eomfs}) with initial conditions $\mathcal{A}_{\text{cl}}=\mathcal{A}_{t_0}$ and $E_{\text{cl}}=E_{t_0}$ at initial time $t_0$.
Ensemble averages at initial time are taken to correspond to the respective quantum expectation values for the gauge fields.
The gauge field dynamics in the classical-statistical approximation is accurately described in the presence of sufficiently high occupation numbers or fields, which is in general the case for the relevant field strengths for pair production. It breaks down once the typical gauge field occupancies become of order unity.
For an introductory review see Ref.~\cite{Berges:2004yj}. 

The subsequent time evolution then follows from (\ref{fmla_th_eomfs}) with 
\begin{equation}
\label{fmla_th_eomsource}
\langle j^\nu(x,t) \rangle = \frac{e}{2}\,\langle{\left[\bar{\psi}(x,t),\gamma^\nu\psi(x,t)\right]}\rangle \ , 
\end{equation}
where the expectation value is taken with respect to the initial state of the spinor field. 
We will restrict ourselves to the Dirac vacuum within the current investigation.
The equations of motion for the spinors read:
\begin{subequations}
\label{fmla_th_eom1}
\begin{align}
 \label{fmla_th_eom_d}
 i\gamma^\mu D_{\mu}\psi(x,t)&=m\psi(x,t) \ , \\
 iD_{\mu}^*\bar{\psi}(x,t)\gamma^\mu&=-m\bar{\psi}(x,t)  \ . 
\end{align}
\end{subequations}
Since the fermions appear only quadratically in the action (\ref{fmla_th_action_cont}) these equations are exact for given classical gauge field configuration. 

Equivalently, the above equations can be conveniently expressed in terms of the equal-time statistical propagator
\begin{equation}
 \label{fmla_th_stat}
 F(x,y;t)\equiv\frac{1}{2}\langle{\left[\psi(x,t),\bar{\psi}(y,t)\right]}\rangle \ ,
\end{equation}
which yields the closed system of equations:
\begin{subequations}
\label{fmla_th_eom2}
\begin{eqnarray}
 i\gamma^\mu D_{x,\mu}F(x,y;t)&=&mF(x,y;t) \ , \\
 iD_{\mu,y}^*F(x,y;t)\gamma^\mu&=&-mF(x,y;t) \ ,\\
 \partial_\mu \mathcal{F}^{\mu\nu}(x,t)&=&-e\tr{\gamma^\nu F(x,x;t)} \ .
\end{eqnarray}
\end{subequations}
We note that the gauge field dynamics in $1+1$ dimensions is special since it is governed by the fermionic backreaction only.
Therefore, we do not consider sampling over initial gauge field configurations in this work.

\subsubsection{Initial conditions and low-cost fermions}

We have to solve the Cauchy problem (\ref{fmla_th_eom2}) in order to calculate fermion--anti-fermion pair production.
Accordingly, we need to provide an initial value for the statistical propagator at $t_0=0$.
To this end, we consider an asymptotic Dirac vacuum -- corresponding to zero particle number and vanishing gauge field -- and employ the framework of canonical quantization:
\begin{equation}
 \psi(x,t)=\int{\frac{dp}{2\pi}e^{ipx}[u(p)a(p)e^{-i\omega t}+v(-p)b^\dagger(-p)e^{i\omega t}]} \ ,
\end{equation}
with $\omega=\sqrt{m^2+p^2}$ and anti-commuting creation and annihilation operators
\begin{equation}
 \{a(p),a^\dagger(p')\}=\{b(p),b^\dagger(p')\}= 2\pi\, \delta(p-p') \ ,
\end{equation}
whereas all other anti-commutators vanish.
An explicit representation of the eigenspinors is given by
\begin{subequations}
\begin{eqnarray}
 u(p)&=&\frac{1}{\sqrt{2\omega(\omega+p)}}\left(\begin{matrix}\omega+p\\m\end{matrix}\right) \ , \\
 v(p)&=&\frac{1}{\sqrt{2\omega(\omega+p)}}\left(\begin{matrix}\omega+p\\-m\end{matrix}\right) \ , 
\end{eqnarray}
\end{subequations}
fulfilling the orthogonality relations:
\begin{equation}
 u^\dagger(p)u(p)=1=v^\dagger(p)v(p) \ \  , \ \  u^\dagger(p)v(-p)=0 \ .
\end{equation}
Because of the fact that the asymptotic Dirac vacuum is homogeneous in space and time, we obtain the initial value:
\begin{equation}
 \label{fmla_th_stat_init}
 F(x,y;t_0)=\int{\frac{dp}{2\pi}e^{ip(x-y)}\frac{m-p\gamma^1}{2\omega}} \ .
\end{equation}

The solution of the time evolution equation of the statistical propagator (\ref{fmla_th_anticomm}) may be based on a mode function expansion \cite{Aarts:1998td}. This treatment can be well suited for low dimensional systems but becomes computationally too expensive in higher dimensions. In view of later applications of our approach to $3+1$ dimensional systems, we perform a stochastic integration of an equivalent set of equations going by the name of low-cost fermions \cite{Borsanyi:2008eu}. To this end, we introduce ensembles of classical stochastic spinors, termed male $\psi_M(x,t)$ and female $\psi_F(x,t)$, instead of spinor field operators.
Given these c-number spinors, we define:
\begin{equation}
 \label{fmla_th_stat_mf}
 F_{\mathrm{sto}}(x,y;t)\equiv\statbracket{\psi_M(x,t)\bar{\psi}_F(y,t)}=\statbracket{\psi_F(x,t)\bar{\psi}_M(y,t)} \ ,
\end{equation}
where here $\langle...\rangle$ is understood as an ensemble average. The requirement
\begin{equation}
 F_{\mathrm{sto}}(x,y;t)\stackrel{!}{=}F(x,y;t) 
\end{equation}
is met provided that the stochastic spinors $\psi_g(x,t)$, with the gender index $g=\{M,F\}$ both satisfy the Dirac equation (\ref{fmla_th_eom_d}) and $F_{\mathrm{sto}}(x,y;t)$ takes the initial value (\ref{fmla_th_stat_init}). 
This second requirement is achieved by initializing the stochastic spinors according to
\begin{equation}
 \psi_g(x,t_0)=\int{\frac{dp}{2\pi}e^{ipx}\frac{1}{\sqrt{2}}[u(p)\xi(p)\pm v(-p)\eta(p)]} \ ,
\end{equation}
with complex random variables $\xi(p)$ and $\eta(p)$.
Note that the male and female spinors only differ by the sign of the antiparticle component.
In order to reproduce the initial value (\ref{fmla_th_stat_init}), the random variables are sampled according to
\begin{equation}
 \statbracket{\xi(p)\xi^*(p')}=\statbracket{\eta(p)\eta^*(p')}=(2\pi)\delta(p-p') \ ,
\end{equation}
whereas all other correlators vanish. 

In an actual simulation employing low-cost fermions, the closed system (\ref{fmla_th_eom2}) is solved in the form:
\begin{subequations}
\label{fmla_th_eom_sto}
\begin{eqnarray}
 i\gamma^\mu D_{\mu}\psi_g(x,t)&=&m\psi_g(x,t) \ , \\
 \label{fmla_th_eom_sto_fs}
 \partial_\mu \mathcal{F}^{\mu\nu}(x,t)&=&-e\tr{\gamma^\nu F_\mathrm{sto}(x,x;t)} \ .
\end{eqnarray}
\end{subequations}
The stochastic spinors $\psi_g(x,t)$ are evolved in time independently and the ensemble average $\langle...\rangle$ appearing in the definition (\ref{fmla_th_stat_mf}) is approximated by an average over a sufficiently large number $N_\mathrm{sto}$ of pairs of male/female spinors. While the computational cost of the mode function approach scales with the volume of the phase space, i.e.\ $N_s^{2d}$ in $d$ dimensions, the resource requirements of 'low-cost' fermions are proportional to just $N_s^d N_\mathrm{sto}$.

\subsection{Lattice formulation}
\label{sec:theory_lattice}

We solve the equations of motion (\ref{fmla_th_eom_sto}) on a $1+1$ dimensional space-time lattice. 
For the spatial sublattice, we define:
\begin{equation}
 \Lambda=\left\{l\,\left|\,\frac{x}{a_s}\right. \in \{0,...,N_s-1\}\right\} \ ,
\end{equation}
with the spatial lattice spacing $a_s$ and the total number of spatial lattice sites $N_s$. A point on the space-time lattice is then denoted by $\mathbf{x}\equiv(l,j)$ with the temporal lattice spacing $a_t$ such that $t=a_t j$.
We employ periodic boundary conditions in the compactified spatial direction whereas no periodicity assumptions apply for the non-compact temporal direction.
The lattice action governing the dynamics is then composed of a pure gauge part as well as part describing the fermions, including their interaction with the gauge field.

\subsubsection{Pure gauge part}

In order to put the gauge fields on the lattice, we use the compact formulation of a gauge theory with $U(1)$ symmetry.
The parallel transporter $U_\mu(\mathbf{x})$ is associated with the link from a lattice point $\mathbf{x}$ to a neighboring point $\mathbf{x}+\hat{\mu}$ in the direction of the space-time lattice axis $\mu=0,1$:
\begin{equation}
 U_\mu(\mathbf{x})=e^{iea_\mu \mathcal{A}_\mu(\mathbf{x})} \ .
\end{equation}
The link variable obeys $U_\mu^*(\mathbf{x})=U_\mu^{-1}(\mathbf{x})$ and we use the definition $U_{-\mu}(\mathbf{x})=U_\mu^*(\mathbf{x}-\hat{\mu})$.
The continuum gauge transformation (\ref{fmla_th_gauge_cont}) translates to
\begin{subequations}
\label{fmla_th_gauge_lat}
\begin{eqnarray}
 \psi(\mathbf{x})&\to&\Omega(\mathbf{x})\psi(\mathbf{x}) \ , \\
 U_\mu(\mathbf{x})&\to&\Omega(\mathbf{x})U_\mu(\mathbf{x})\Omega^*(\mathbf{x}+\hat{\mu}) \ ,
\end{eqnarray}
\end{subequations}
with $\Omega\in U(1)$. Given the gauge-dependent link variable, we define the gauge-invariant plaquette variable:
\begin{equation}
 \label{fmla_th_plaq}
 U_{\mu\nu}(\mathbf{x})=U_\mu(\mathbf{x})U_\nu(\mathbf{x}+\hat{\mu})U^*_\mu(\mathbf{x}+\hat{\nu})U^*_\nu(\mathbf{x}) \ .
\end{equation}
Disregarding higher order terms in the lattice spacings $a_\mu$, we find:
\begin{equation}
 \label{fmla_th_temp_plaq}
 U_{\mu\nu}(\mathbf{x})=e^{iea_\mu a_\nu \mathcal{F}_{\mu\nu}(\mathbf{x})} \ .
\end{equation}
Accordingly, the pure gauge part of the action can be written as
\begin{equation}
 \mathcal{S}_g[U]=\frac{1}{e^2a_sa_t}\sum_{\mathbf{x}}{\re{1-U_{01}(\mathbf{x})}} \ .
\end{equation}
Moreover, the electric field (\ref{fmla_th_el_cont}) is given by
\begin{equation}
 E(\mathbf{x})=\frac{1}{ea_sa_t}\im{U_{01}(\mathbf{x})} \ . 
\end{equation}

\subsubsection{Dirac and interaction part}

Using a symmetric finite difference approximation for the first derivatives, the naive discretization of the fermionic part is given by
\begin{align}
\mathcal{S}&_f^{(0)}[\psi,\bar{\psi},U]= a_ta_s\sum_{\mathbf{x}}\bar{\psi}(\mathbf{x}) \nonumber \\
& \times \left[i\gamma^\mu\frac{U_\mu(\mathbf{x})\psi(\mathbf{x}+\hat{\mu})-U_{-\mu}(\mathbf{x})\psi(\mathbf{x}-\hat{\mu})}{2a_\mu}-m\psi(\mathbf{x})\right] 
\end{align}
where the gender index is omitted for simplicity.
This expression is gauge-invariant under lattice gauge transformations (\ref{fmla_th_gauge_lat}), however, it also gives rise to unphysical states.
This fermion doubling problem is well-known from Euclidean lattice field theory.
However, unlike to Euclidean time one distinguishes between temporal and spatial doubler modes in the real-time formulation \cite{Aarts:1998td}.

The spatial doublers, corresponding to high-momentum excitations showing a low-energy dispersion relation, are conveniently suppressed by adding a higher derivative term to the action:
\begin{equation}
 -\frac{a_s}{2}\int{d^2x}\,\bar{\psi}D_1D^1\psi \ ,
\end{equation}
which vanishes in the continuum limit $a_s\to0$.
This gauge-invariant Wilson term in space ensures that only low-momentum excitations show a low-energy dispersion relation.
In the lattice implementation, this corresponds to adding one more term to the action:
\begin{align}
\mathcal{S}&_f^{(W)}[\psi,\bar{\psi},U]= a_ta_s\sum_{\mathbf{x}} \bar{\psi}(\mathbf{x}) \nonumber \\
& \times \bigg[\frac{U_1(\mathbf{x})\psi(\mathbf{x}+\hat{1})-2\psi(\mathbf{x})+U_{-1}(\mathbf{x})\psi(\mathbf{x}-\hat{1})}{2a_s}\bigg] \ .
\end{align}
We do not include a temporal Wilson term as this would turn the Dirac equation into a second order differential equation in time. 
The temporal doublers are avoided provided that we initialize only the physical mode and choose the temporal lattice spacing to be much smaller than the spatial lattice spacing $a_t\ll a_s$ \cite{Aarts:1998td,Borsanyi:2008eu,Saffin:2011kc}.

\subsubsection{Low-cost Wilson fermions}

The construction of the stochastic spinor ensemble on the space-time lattice follows the same lines as in the continuum outlined in Sec.~\ref{sec:theory_low}.
For spatial momenta $p$ we define the conjugate lattice:
\begin{equation}
 \tilde{\Lambda}=\left\{ q \, \left| \, \frac{L p}{2\pi} \right. \in \left\{-\frac{N_s}{2},...,\frac{N_s}{2}-1\right\}\right\} \ ,
\end{equation}
with the spatial volume $L=a_sN_s$.
Accordingly, the discrete Fourier transformation is given by
\begin{subequations}
\begin{eqnarray}
 \psi(\mathbf{x})&=&\frac{1}{L}\sum_{q\in\tilde{\Lambda}}\exp\left(\frac{2\pi ilq}{N_s}\right)\tilde{\psi}(\mathbf{q}) \ , \\
 \tilde{\psi}(\mathbf{q})&=&a_s\sum_{l\in\Lambda}\exp\left(-\frac{2\pi ilq}{N_s}\right)\psi(\mathbf{x}) \ ,
\end{eqnarray}
\end{subequations}
with the notation $\mathbf{q}\equiv(q,j)$. The stochastic spinors are then again initialized at $t_0=a_tj_0=0$:
\begin{equation}
 \tilde{\psi}_g(\mathbf{q}_0)=\frac{1}{\sqrt{2}}[u(q)\xi(q)\pm v(-q)\eta(q)] \ ,
\end{equation}
with $\mathbf{q}_0=(q,j_0)$ and eigenspinors:
\begin{subequations}
\begin{eqnarray}
 u(q)&=&\frac{1}{\sqrt{2\tilde{\omega}(\tilde{\omega}+\tilde{q})}}\left(\begin{matrix}\tilde{\omega}+\tilde{q}\\\tilde{m}\end{matrix}\right) \ , \\
 v(q)&=&\frac{1}{\sqrt{2\tilde{\omega}(\tilde{\omega}+\tilde{q})}}\left(\begin{matrix}\tilde{\omega}+\tilde{q}\\-\tilde{m}\end{matrix}\right) \ .
\end{eqnarray}
\end{subequations}
Here the mass term is modified due to the spatial Wilson term
\begin{equation}
 \tilde{m}=m+\frac{2}{a_s}\sin^2\left(\frac{\pi q}{N_s}\right) \ ,
\end{equation}
and we define the lattice quantities
\begin{equation}
 \tilde{q}=\frac{1}{a_s}\sin\left(\frac{2\pi q}{N_s}\right) \quad , \quad \tilde{\omega}=\sqrt{\tilde{m}^2+\tilde{q}^2} \ .
\end{equation}

In order to reproduce the correct initial value for the statistical propagator on the lattice,
\begin{equation}
 F(l_1,l_2;j_0)=\frac{1}{L}\sum_{q\in\tilde{\Lambda}}\exp\left(\frac{2\pi i(l_1-l_2)q}{N_s}\right)\frac{\tilde{m}-\tilde{q}\gamma^1}{2\tilde{\omega}} \ ,
\end{equation}
the complex random variables $\xi(q)$ and $\eta(q)$ are sampled according to
\begin{equation}
  \statbracket{\xi(q)\xi^*(q')}=\statbracket{\eta(q)\eta^*(q')}=L\delta_{q,q'} \ .
\end{equation}
This is most easily done by assuming
\begin{equation}
 \xi(q)=X(q)e^{i\phi(q)} \quad , \quad \eta(q)=Y(q)e^{i\theta(q)} 
\end{equation}
and choosing the amplitudes $X(q)$ and $Y(q)$ to be Gaussian distributed whereas the phases $\phi(q)$ and $\theta(q)$ are chosen to be uniformly distributed on the interval $[-\pi,\pi)$.
\newline
\subsubsection{Lattice equations of motion}

To simplify simulations afterwards, we use the gauge freedom and employ the lattice equivalent of the temporal axial gauge: $U_0(\mathbf{x})=1$ for the equations of motion. Stationarity of the lattice action 
\begin{equation}
\label{fmla_th_action_lat} \mathcal{S}[\psi,\bar{\psi},U]=\mathcal{S}_g[U]+\mathcal{S}_f^{(0)}[\psi,\bar{\psi},U]+\mathcal{S}_f^{(W)}[\psi,\bar{\psi},U] 
\end{equation}
\newline
with respect to the temporal link $U_0(\mathbf{x})$ results in the discretized version of the Gauss law:
\begin{equation}
 \label{fmla_th_eom_lat1}
 E(\mathbf{x})-E(\mathbf{x}-\hat{1})=\frac{ea_s}{2}\bar{\psi}(\mathbf{x})\gamma^0\psi(\mathbf{x}+\hat{0})+c.c. 
\end{equation}
This equation is a constraint which is fulfilled during the time evolution for the considered initial conditions.

The stationary condition of the action with respect to the spatial link $U_1(\mathbf{x})$, on the other hand, results in the equation of motion:
\begin{align}
 \label{fmla_th_eom_lat2}
 E(\mathbf{x})-&E(\mathbf{x}-\hat{0})= \nonumber \\
              &-\frac{ea_t}{2}\bar{\psi}(\mathbf{x})[\gamma^1-i]U_1(\mathbf{x})\psi(\mathbf{x}+\hat{1})+c.c.
\end{align}

Finally, the stationarity condition of the action with respect to the Dirac field $\bar{\psi}(\mathbf{x})$  gives:
\begin{widetext}
\begin{eqnarray}
 \label{fmla_th_eom_lat3}
 \psi(\mathbf{x}+\hat{0})=\psi(\mathbf{x}-\hat{0})-2ia_t\left(m+\frac{1}{a_s}\right)\gamma^0\psi(\mathbf{x})-\frac{a_t}{a_s}\left(\gamma^0[\gamma^1-i]U_1(\mathbf{x})\psi(\mathbf{x}+\hat{1})-\gamma^0[\gamma^1+i]U_{-1}(\mathbf{x})\psi(\mathbf{x}-\hat{1})\right) \ .
\end{eqnarray}
\end{widetext}
The set of equations (\ref{fmla_th_eom_lat1})--(\ref{fmla_th_eom_lat3}) is the lattice version of (\ref{fmla_th_eom_sto}) in temporal axial gauge including a spatial Wilson term.

In order to solve the Cauchy problem, we have to provide the following initial values at $t_0=a_t j_0=0$:
\begin{equation}
 E(\mathbf{x}_0-\hat{0}) \quad , \quad U_1(\mathbf{x}_0) \quad , \quad \psi(\mathbf{x}_0-\hat{0}) \quad , \quad \psi(\mathbf{x}_0)  \nonumber
\end{equation} 
with $\mathbf{x}_0=(l,j_0)$ for all $l\in\Lambda$.
Most notably, we have to choose initial values for the spinors at $j_0-1$ and $j_0$, which is a consequence of the chosen leapfrog algorithm. To be able to initialize them we assume a free field evolution at initial times.

The algorithm, which is a variant of the one introduced in \cite{Ambjorn:1990pu}, can then be summarized in the following way:
\begin{itemize}
 \item[1.]{{\it{Electric field evolution:}} Given $E(\mathbf{x}-\hat{0})$, $U_1(\mathbf{x})$ and $\psi(\mathbf{x})$ we evolve the electric field to $E(\mathbf{x})$ according to (\ref{fmla_th_eom_lat2}).}
 \item[2.]{{\it{Dirac field evolution:}} Given $\psi(\mathbf{x}-\hat{0})$, $\psi(\mathbf{x})$ and $U_1(\mathbf{x})$ we evolve the Dirac field to $\psi(\mathbf{x}+\hat{0})$ according to (\ref{fmla_th_eom_lat3}).}
 \item[3.]{{\it{Temporal plaquette:}} We evaluate the temporal plaquette $U_{01}(\mathbf{x})$ according to (\ref{fmla_th_temp_plaq}):
 \begin{equation}
  U_{01}(\mathbf{x})=e^{iea_sa_t\mathcal{F}_{01}(\mathbf{x})}=e^{iea_sa_tE(\mathbf{x})} \ .
 \end{equation}}
 \item[4.]{{\it{Spatial link evolution:}} The link variable $U_1(\mathbf{x}+\hat{0})$ is calculated from the temporal plaquette $U_{01}(\mathbf{x})$ in temporal axial gauge according to (\ref{fmla_th_plaq}): 
 \begin{equation}
  U_1(\mathbf{x}+\hat{0})=U_{01}(\mathbf{x})U_1(\mathbf{x}) \ .
 \end{equation}}
 \item[5.]{Reiterate the steps $1$ -- $4$.}
\end{itemize}

\subsection{Gauge-invariant correlation functions}
\label{sec:theory_DHW}

In order to compare our simulation results with typical discussions using the Dirac-Heisenberg-Wigner phase-space approach \cite{BialynickiBirula:1991tx,Hebenstreit:2010vz,Skokov:2007gy,Hebenstreit:2010cc,BialynickiBirula:2011uh,Hebenstreit:2011wk,Hebenstreit:2011cr}, we define suitable gauge invariant two-point correlation functions on the lattice.

\subsubsection{Continuum Wigner function}

Starting from the continuum expression for the statistical propagator (\ref{fmla_th_stat}), a gauge-invariant generalization may be defined as:
\begin{equation}
 \label{fmla_th_stat_g}
 \tilde{F}(x_1,x_2;t)=\exp\left(ie\int_{x_2}^{x_1}{dx\mathcal{A}(x,t)}\right)F(x_1,x_2;t) \ .
\end{equation}
The Wilson line factor ensures gauge invariance under local $U(1)$ transformations.
The Fourier transformation with respect to the relative coordinate defines the Wigner function:
\begin{equation}
 \label{fmla_th_wig_cont1}
 \mathcal{W}(x,p,t)\equiv-\int{dye^{-ipy}\tilde{F}\left(x+y/2,x-y/2;t\right)} \ ,
\end{equation}
with $x=(x_1+x_2)/2$ and $y=x_1-x_2$.
The Wilson line factor in (\ref{fmla_th_stat_g}) is not unique, however, a physical sensible interpretation of $p$ as kinetic momentum forces the integration path to be chosen along the straight line.
Equivalently to (\ref{fmla_th_wig_cont1}), we may also write:
\begin{equation}
 \label{fmla_th_wig_cont2}
 \mathcal{W}(x,p,t)=-\int{dze^{2ip(x-z)}\tilde{F}(z,2x-z;t)}+\gamma.c.
\end{equation}
with the abbreviation:
\begin{equation}
 D+\gamma.c.\equiv D+\gamma^0D^\dagger\gamma^0 \ .
\end{equation}
As the Wigner function is in the Dirac algebra and fulfills $\mathcal{W}^\dagger=\gamma^0\mathcal{W}\gamma^0$, one can decompose it in terms of its Dirac bilinears:
\begin{equation}
 \label{fmla_th_wig_irred}
 \mathcal{W}=\frac{1}{2}\left[\mathbbm{s}+i\gamma_5\mathbbm{p}+\gamma^0\mathbbm{v}_0-\gamma^1\mathbbm{v}\right] \ , 
\end{equation}
where all its irreducible components can be chosen to be real. 
Regarding the Dirac vacuum, which is described by the statistical propagator (\ref{fmla_th_stat_init}), the only non-vanishing components are given by:
\begin{equation}
 \mathbbm{s}_\mathrm{vac}(x,p,t)=-\frac{m}{\omega} \quad , \quad \mathbbm{v}_\mathrm{vac}(x,p,t)=-\frac{p}{\omega} \ .
\end{equation}

In terms of these components the total charge $\mathcal{Q}$ and the total energy $\mathcal{E}$ can be expressed as phase-space integrals:
\begin{subequations}
\begin{align}
 \mathcal{Q}&=e\int{d\Gamma \mathbbm{v}_0(x,p,t)} \ , \\
 \label{fmla_th_energy}
 \mathcal{E}&=\int{d\Gamma [m\mathbbm{s}(x,p,t)+p\mathbbm{v}(x,p,t)]}+\frac{1}{2}\int{dxE^2(x,t)} \ ,
\end{align}
\end{subequations}
with the phase-space volume element $d\Gamma=dxdp/(2\pi)$.
The integrands $\epsilon(x,p,t)=[m\mathbbm{s}(x,p,t)+p\mathbbm{v}(x,p,t)]$ and $\varrho(x,p,t)=\mathbbm{v}_0(x,p,t)$ are regarded as energy pseudo-distribution and charge pseudo-distribution, respectively.
We may define further quantities such as the particle number pseudo-distributions:
\begin{equation}
 n^{\pm}(x,p,t)=\frac{\epsilon(x,p,t)-\epsilon_\mathrm{vac}(x,p,t)\pm\omega\mathbbm{v}_0(x,p,t)}{2\omega} \ ,
\end{equation}
which may be associated to the density of particles and anti-particles, respectively.
Of course, in the interacting quantum theory the interpretation of these phase-space pseudo-distributions, collectively denoted as $m(x,p,t)$, has to be taken with care. We emphasize that our approach is not based on these quantities and we use them only for read-out and comparison with literature results. We will frequently consider also the partially integrated position space and momentum space marginal distributions:
\begin{subequations}
\begin{eqnarray}
 m_\mathcal{X}(x,t)&\equiv&\int{\frac{dp}{2\pi}m(x,p,t)} \ , \\
 m_\mathcal{P}(p,t)&\equiv&\int{dx\, m(x,p,t)} \ ,
\end{eqnarray}
\end{subequations}
or the fully integrated quantities:
\begin{equation}
 m(t)\equiv\int{d\Gamma\, m(x,p,t)} \ ,
\end{equation}
instead of the pseudo-distributions $m(x,p,t)$.

\subsubsection{Lattice Wigner function}

In order to adjust the above continuum treatment to the lattice, we have to account for the periodicity of the spatial lattice properly.
Our approach is an extension of previous work on the discrete Wigner function in the context of signal processing \cite{Peyrin:1986}.

We first define the gauge invariant generalization of the lattice statistical propagator according to:
\begin{equation}
 \tilde{F}(l_1,l_2;j)=\mathcal{U}(l_1,l_2;j)F(l_1,l_2;j) \ ,
\end{equation}
where $\mathcal{U}(l_1,l_2;j)$ is the lattice analogue of the Wilson line factor along the straight line path.
However, since the straight line path between two lattice points is not unique due the periodicity of the lattice, we choose to define it such that properties of the above standard continuum interpretation apply. It turns out that this requires taking the shortest path between two lattice points.
Accordingly, for $\Delta l=l_1-l_2>0$ we employ:
\begin{subequations}
\begin{eqnarray}
 &\Delta l\leq \displaystyle \frac{N_s}{2}: \quad &\mathcal{U}=\prod\limits_{l=l_2}^{l_1-1}{U_1(\mathbf{x})} \ ,\qquad \\
 &\Delta l> \displaystyle \frac{N_s}{2}: \quad &\mathcal{U}=\prod\limits_{l=l_1}^{N_s-1}{U_1^*(\mathbf{x})}\times\prod\limits_{l=0}^{l_2-1}{U_1^*(\mathbf{x})} \ . \qquad
\end{eqnarray}
\end{subequations}
On the other hand, for $\Delta l<0$ we use:
\begin{subequations}
\begin{eqnarray}
 &\Delta l> -\frac{N_s}{2}: \quad &\mathcal{U}=\prod\limits_{l=l_1}^{l_2-1}{U_1^*(\mathbf{x})} \ ,\qquad \\
 &\Delta l\leq -\frac{N_s}{2}: \quad &\mathcal{U}=\prod\limits_{l=l_2}^{N_s-1}{U_1(\mathbf{x})}\times\prod\limits_{l=0}^{l_1-1}{U_1(\mathbf{x})} \ . \qquad
\end{eqnarray}
\end{subequations}

More precisely, we utilize the following Wigner lattices:
\begin{subequations}
\begin{eqnarray}
 \Lambda_\mathcal{W}&=&\left\{l \, \left| \, \frac{2 x}{a_s} \right. \in \left\{0,...,2N_s-1\right\}\right\} \ , \\
 \tilde{\Lambda}_\mathcal{W}&=&\left\{q \, \left| \, \frac{L p}{\pi} \right. \in \left\{-N_s,...,N_s-1\right\}\right\} \ ,
\end{eqnarray}
\end{subequations}
which have the same extent as the original ones $\Lambda$ and $\tilde{\Lambda}$, however, each with twice as many grid points.
We then define the lattice Wigner function according to
\begin{align}
 \label{fmla_th_wig_lat}
 \mathcal{W}(l,q,j)\equiv-&\frac{a_s}{2}e^{\pi i l q/N_s} \nonumber \\ 
 &\times\sum_{k\in\Lambda}{e^{-2\pi i k q/N_s}\tilde{F}(k,[l-k]_{N_s};j)} + \gamma.c.
\end{align}
with $l\in\Lambda_\mathcal{W}$ and $q\in\tilde{\Lambda}_\mathcal{W}$.
We account for the periodicity of the lattice by taking the module operation in the second argument of the statistical propagator:
\begin{equation}
 [l-k]_{N_s}=(l-k)\operatorname{mod}N_s \ .
\end{equation}
This definition is such that we reproduce the above continuum expressions for the marginal distributions, as shown in Appendix \ref{ap:marg}.
Moreover, the lattice Wigner function (\ref{fmla_th_wig_lat}) again fulfills $\mathcal{W}^\dagger=\gamma^0\mathcal{W}\gamma^0$ so that the decomposition in terms of its Dirac bilinears (\ref{fmla_th_wig_irred}) is possible.

In complete analogy to the continuum, we may then again define various pseudo distributions:
\begin{subequations}
\label{fmla_th_pseud_lat}
\begin{eqnarray}
 \varrho(l,q,t)&=&e\mathbbm{v}_0(l,q,t) \, , \\
 \epsilon(l,q,t)&=&[\tilde{m}\mathbbm{s}(l,q,t)+\tilde{q}\mathbbm{v}(l,q,t)] \, , \\
 n^\pm(l,q,t)&=&\frac{\epsilon(l,q,t)-\epsilon_\mathrm{vac}(l,q,t)\pm\tilde{\omega}\mathbbm{v}_0(l,q,t)}{2\tilde{\omega}} \, , \qquad
\end{eqnarray}
\end{subequations}
corresponding to charge, energy and particle/anti-particle number, respectively.
Given these pseudo-distributions $m(l,q,t)$, the marginal distributions are defined via
\begin{subequations}
\begin{eqnarray}
 m_\mathcal{X}(l,j)&\equiv&\frac{1}{2L}\sum_{q\in\tilde\Lambda_\mathcal{W}}{m(l,q,j)} \ , \\
 m_\mathcal{P}(q,j)&\equiv&\frac{a_s}{2}\sum_{l\in\Lambda_\mathcal{W}}{m(l,q,j)} \ ,
\end{eqnarray}
\end{subequations}
whereas the fully integrated quantities are given by
\begin{equation}
 m(j)=\frac{1}{2N_s}\sum_{q\in\tilde{\Lambda}}\sum_{l\in\Lambda_\mathcal{W}}m(l,q,j) \ .
\end{equation}
Here one should note the summation order in the last expression: The sum over $l\in\Lambda_\mathcal{W}$ yields the marginal distribution $m_\mathcal{P}(q,t)$ which is non-vanishing for even $q$ only. 
Accordingly, the subsequent sum is just taken over $q\in\tilde{\Lambda}$.


\section{Pair production simulations}
\label{sec:results}

We now come to the results which are based on the lattice approach presented in the previous section.
As a first example, we consider a static electric background field, disregarding the backreaction of created fermion--anti-fermion pairs.
This configuration can be solved analytically such that we can compare our lattice simulations with well established continuum results.
Subsequently, we also include the backreaction of created fermion--anti-fermion pairs and discuss the decay of the gauge field which shuts pair production off after a characteristic time.

As a second example, we investigate the pair creation process in the presence of a space- and time-dependent electric field. Neglecting backreaction in a first step, we can compare to and complement previous investigations based on the continuum Dirac-Heisenberg-Wigner approach \cite{Hebenstreit:2011wk,Hebenstreit:2011pm}. Subsequently, we solve the full lattice evolution and compare.

\subsection{Spatially homogeneous gauge field}

We consider a static electric background field $E(x,t)=E_0$ in temporal axial gauge $\mathcal{A}_0=0$, represented by the vector potential
\begin{equation}
 \mathcal{A}(t)=E_0 t \ . 
\end{equation}
Within the compact lattice formulation, this corresponds to a trivial temporal link $U_0(\mathbf{x})=1$ and the spatial link
\begin{equation}
 U_1(\mathbf{x})=e^{iea_ta_sE_0j} 
\end{equation}
disregarding higher order terms in the lattice spacing.
Moreover, we introduce the dimensionless field strength parameter
\begin{equation}
 \epsilon=\frac{E_0}{E_c} \ ,
\end{equation}
with the critical Schwinger field strength $E_c=m^2/e$. For all subsequent numerical results we employ $e/m = 0.3$. In Appendix \ref{ap:stat} we briefly review some analytic results, which are used for comparison in the following.

\subsubsection{Particle production without backreaction}

\begin{figure}[b]
 \centering
 \includegraphics[width=\columnwidth]{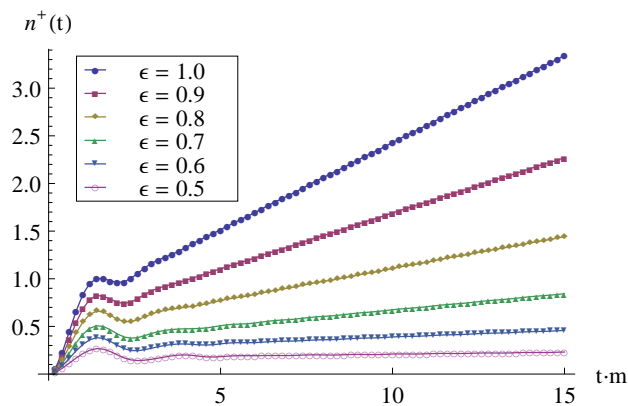}
 \caption{Time evolution of the total particle number $n^+(t)$ for different values of $\epsilon$.
 The parameters are $N_\mathrm{sto}=10^3$, $a_t=0.00125/m$, $a_s=0.025/m$, $N_s=1024$ such that $L=25.6/m$.}
 \label{fig:number_free_static}
\end{figure}

In this section we disregard the backreaction of created fermion--anti-fermion pairs on the electric field. This corresponds to neglecting the fermionic contributions in the gauge field equation of motion (\ref{fmla_th_eom_lat2}). Starting with the vacuum initial conditions for the fermions, this amounts to evolving the fermion equation (\ref{fmla_th_eom_lat3}) with a sudden switching-on of the electric field at initial time. 

In Fig.~\ref{fig:number_free_static} we show the time evolution of the total number of produced particles, $n^+(t)$, for various values of the dimensionless field strength parameter $\epsilon$. Most notably, we observe two different regimes: At early times there is a transient oscillatory behavior superimposed which can be attributed to the sudden switching-on of the electric field. For $\epsilon = 1$ we estimate this oscillation to be exponentially damped with a characteristic rate $\gamma \simeq 1/m$, leading to a purely linear growth to very good accuracy after times of a few $\gamma^{-1}$. 

\begin{figure}[b]
 \centering
 \includegraphics[width=\columnwidth]{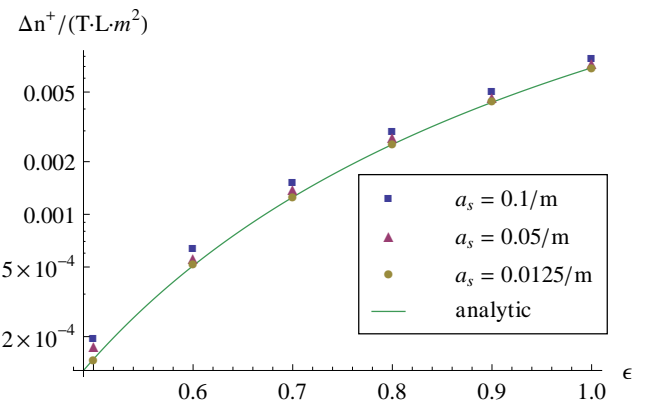}
 \caption{Comparison of the analytical results (\ref{fmla_res_schwinger}) with the numerical fit for $L=25.6/m$ and different lattice spacings $a_s$.
 The remaining parameters are $N_\mathrm{sto}=10^3$ and $a_t=a_s/20$.}
 \label{fig:schwinger_free_static}
\end{figure}

The slope of the linear rise of $n^+(t)$ strongly depends on the value of $\epsilon$. In order to extract its functional dependence, we perform a linear fit. For this we measure the change in the total number of particles $\Delta n^+$ which are produced during the time interval $T=10/m$ for times large compared to $\gamma^{-1}$. In Fig.~\ref{fig:schwinger_free_static} we compare the slope for different values of $\epsilon$ with the analytical result from Appendix \ref{ap:stat}:
\begin{equation}
 \label{fmla_res_schwinger}
 \frac{\Delta n^+}{TLm^2}=\frac{\epsilon}{2\pi}\exp\left(-\frac{\pi}{\epsilon}\right) \ . 
\end{equation}
We emphasize that for this analytical result the initial time is sent to the remote past such that it cannot reproduce the transient oscillatory regime. However, both the simulation and the analytical result should accurately agree for large enough times. 

The lattice results are shown in Fig.~\ref{fig:schwinger_free_static} for different spatial lattice spacings $a_s$ keeping the volume $L=a_sN_s$ constant, thus increasing $N_s$ accordingly. One clearly observes that the simulation and the analytical result (\ref{fmla_res_schwinger}) fall nearly on top of each other for small enough $a_s$, indicating that we are close to the continuum limit in that case. As a matter of fact, we find that temporal discretization errors are quite negligible for $a_t\lesssim a_s/20$.
This corroborates that the real-time lattice simulation is in fact capable of reproducing the analytic results in the continuum limit to very good accuracy.

\begin{figure}[t]
 \centering
 \includegraphics[width=\columnwidth]{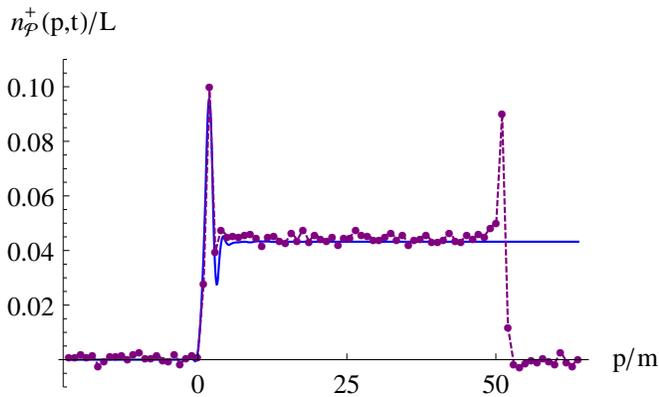}
 \caption{Comparison of the continuum expression $f(p)$ (solid line) with the normalized particle number marginal distribution $n^+_\mathcal{P}(p,t)/L$ (dashed line) for $\epsilon=1$ at $t=50/m$. 
 The parameters are $N_\mathrm{sto}=10^5$, $a_t=0.00125/m$, $a_s=0.025/m$, $N_s=1024$ such that $L=25.6/m$.}
 \label{fig:dist_free_static}
\end{figure}

In Fig.~\ref{fig:dist_free_static} we show the normalized particle number marginal distribution $n^+_\mathcal{P}(p,t)/L$, corresponding to the momentum spectrum of created particles (\ref{fmla_ap_anal}), and compare it to the continuum value $f(p)$.
In comparison to the integrated particle number shown above, the spectrum is not smooth but shows fluctuations due to the sampling of low-cost fermions.
As a matter of fact, these fluctuations can be systematically reduced by taking $N_\mathrm{sto}$ larger. We find that it suffices to take the number $N_\mathrm{sto}$ of the order of $10^3$ in order to accurately calculate integrated quantities such as $n^+(t)$. This is in contrast to the momentum spectrum $n^+_\mathcal{P}(p,t)$ where the number $N_\mathrm{sto}$ needed to be at least of the order of $10^4$ to suppress the statistical fluctuations sufficiently and obtain sensible results. In contrast to the one-dimensional case considered here, the convergence is expected to be even better for three space dimensions where self-averaging plays a major role \cite{Berges:2010zv}.

The established interpretation of $f(p)$ is such that electric field energy is taken and transformed into virtual fermion--anti-fermion pairs, showing up as the distinctive peak around momenta $p=0$.
If the applied field strength $E_0$ is large enough, i.e.\ of the order of $E_c$, these charged excitations can be separated over the Compton wavelength and become real fermion--anti-fermion pairs. 
These real particles are then further accelerated in the background electric field and achieve higher and higher momenta up to $p\to\infty$.

We observe good agreement of simulation and analytical results regarding the virtual fermion--anti-fermion peak around $p=0$ as well as the overall magnitude of $n^+_\mathcal{P}(p,t)/L$. 
However, we observe a qualitatively different behavior for large momenta.
This is due to the fact that the analytic result assumes an electric field which has existed for all times such that all momenta up to $p\to\infty$ are already occupied whereas we solve an initial value problem on the lattice.
Accordingly, we observe a transient effect corresponding to the peak at high momenta propagating to higher and higher momenta during the time evolution.

\subsubsection{Particle production with backreaction}

We now include the backreaction of created fermion--anti-fermion pairs on the electric field.
As a consequence, particle creation comes with a simultaneous decrease of the electric field due to energy conservation.
This energy transfer from the gauge sector to the fermion sector finally results in a decay of the electric field.

\begin{figure}[t]
 \centering
 \includegraphics[width=0.97\columnwidth]{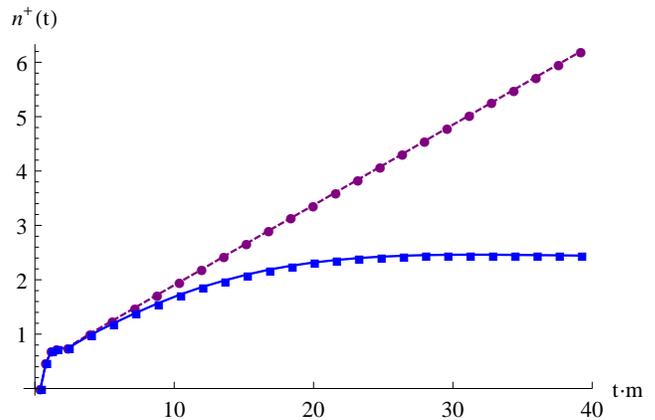}
 \caption{Time evolution of the total particle number $n^+(t)$ with (solid line) and without (dashed line) backreaction for an initial value $\epsilon=1$.
 The parameters are $N_\mathrm{sto}=10^4$, $a_t=0.0025/m$, $a_s=0.075/m$, $N_s=256$ such that $L=19.2/m$.}
 \label{fig:number_back_static}
\end{figure}

In Fig.~\ref{fig:number_back_static} we compare the time evolution of the total particle number $n^+(t)$ for simulations with and without backreaction.
We have already seen in the previous section that the particle number grows eventually linearly if we disregard the backreaction of created fermion--anti-fermion pairs on the electric field.
However, this changes drastically if we include the backreaction mechanism:
Following the transient regime at early times, the pair production rate immediately slows down once the pair creation process kicks in and the electric field is weakened.
Eventually, this is getting to a point where the fermion--anti-fermion production process effectively stops and $n^+(t)$ levels off. This process happens on rather short time scales of the order of $\Delta t\sim25/m$.

\begin{figure}[t]
 \centering
 \includegraphics[width=\columnwidth]{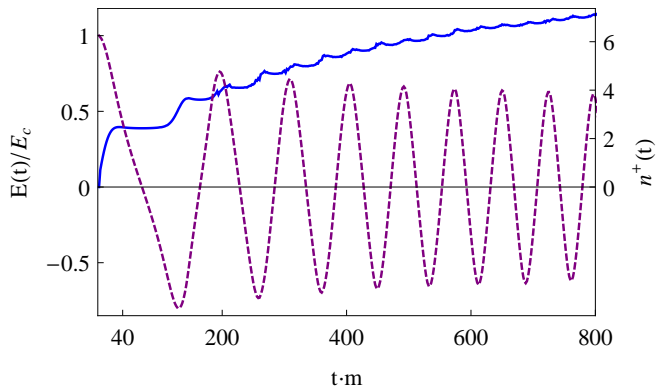}
 \caption{Time evolution of the electric field $E(t)$ (dashed line) and the total particle number $n^+(t)$ (solid line) for an initial value $\epsilon=1$ for a much longer time period.
 The parameters are as in Fig.~\ref{fig:number_back_static}.}
 \label{fig:efld_back_static}
\end{figure}

To see the long-time behavior, in Fig.~\ref{fig:efld_back_static} we show the particle number $n^+(t)$ and the electric field $E(t)$ for times up to $800/m$.
Most notably, we observe the occurrence of plasma oscillations in accordance with previous investigations \cite{Kluger:1992gb}: 
Starting from $t_0=0$, the magnitude of the electric field decreases due to the creation of fermion--anti-fermion pairs.
Due to the backreaction mechanism, an internal electric field builds up so that the field eventually changes sign and grows until a first local minimum is achieved.
The electric field then increases again, changes sign, reaches a local maximum and so forth. 
The oscillation frequency $\Omega$ increases with the number of produced fermions, in accordance with the expected parametric dependence. 

The behavior of the particle number $n^+(t)$ follows from the oscillatory behavior of the electric field: 
Particle creation effectively terminates when the magnitude of the field strength drops below $\sim0.5E_\mathrm{c}$, corresponding to the approximate plateaus in $n^+(t)$.
However, at those instants of time at which the electric field reaches local extrema, fermions are created again.
Due to the fact that the envelope of the electric field decreases with time, the particle number $n^+(t)$ assumes the shape of a staircase with decreasing step height. 

We emphasize that the classicality condition $\langle \mathcal{A} \mathcal{A} \rangle \gg 1$ \cite{QvsCS1} is well fulfilled also after the backreaction effectively terminates the pair production: For an electric field amplitude $E$ with characteristic oscillation frequency $\Omega$ the classicality condition reads $E^2/\Omega^2 \gg 1$. In our case $E\simeq E_c/2 = m^2/2e$ during these times such that with $\Omega \simeq  \pi m/50$ for the employed coupling $e/m = 0.3$ we have $E^2/\Omega^2 \simeq 700$. 

\begin{figure}[b]
 \centering
 \includegraphics[width=0.97\columnwidth]{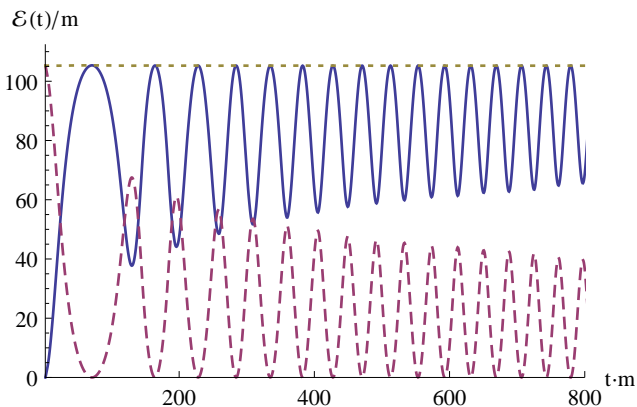}
 \caption{Energy transfer between the fermionic sector (solid line) and the gauge sector (dashed line) for an initial value \mbox{$\epsilon=1$}. The dotted line shows the total energy, with the fermion vacuum contribution being subtracted. The parameters are as in Fig.~\ref{fig:number_back_static}.}
 \label{fig:energy_back_static}
\end{figure}

Moreover, in Fig.~\ref{fig:energy_back_static} we demonstrate that the energy transfer from the gauge sector to the fermion sector is in agreement with energy conservation.

\begin{figure}[b]
 \centering
 \includegraphics[width=0.97\columnwidth]{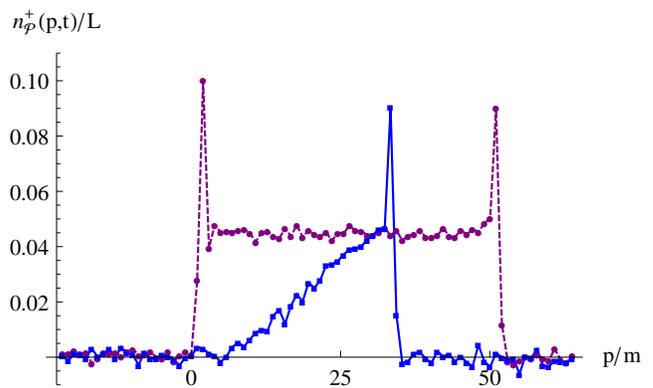}
 \caption{Normalized particle number marginal distribution $n^+_\mathcal{P}(p,t)/L$ with (solid line) and without (dashed line) backreaction for an initial value $\epsilon=1$ at $t=50/m$. 
 The parameters are $N_\mathrm{sto}=10^5$, $a_t=0.00125/m$, $a_s=0.025/m$, $N_s=1024$ such that $L=25.6/m$.}
 \label{fig:dist_back_static}
\end{figure}

Finally, in Fig.~\ref{fig:dist_back_static} we compare the normalized particle number marginal distributions $n^+_\mathcal{P}(p,t)/L$ for simulations with and without backreaction prior to the onset of plasma oscillations.
We observe two major modifications if we include the backreaction:

First, the high-momentum peak is shifted to lower momenta.
This is due to the fact that acceleration in an electric field is proportional to its field strength.
Accordingly, particles are less accelerated and achieve lower momenta if the electric field is decreasing gradually.

Second, the overall magnitude of $n^+_\mathcal{P}(p,t)/L$ declines in the low-momentum regime.
Again, this can be attributed to the decay of the electric field as the decrease of the field strength is accompanied by a drop in the pair production rate.
Consequently, this gradual decrease of the pair production rate shows itself as a decreasing amplitude of $n^+_\mathcal{P}(p,t)/L$.

\begin{figure}[b]
 \centering
 \includegraphics[width=\columnwidth]{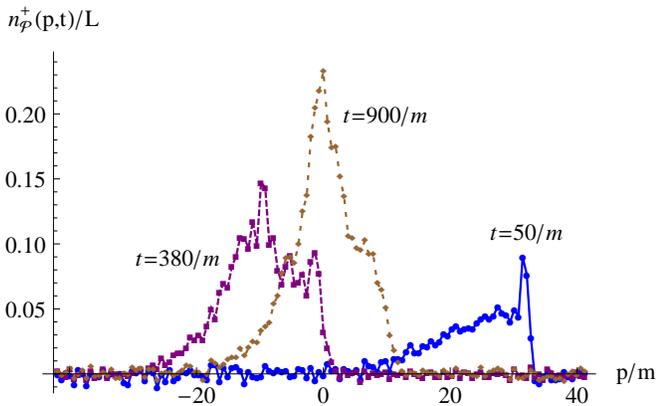}
 \caption{Normalized particle number marginal distribution $n^+_\mathcal{P}(p,t)/L$ for simulations with backreaction at different times for an initial value $\epsilon=1$. 
 The parameters are $N_\mathrm{sto}=10^4$, $a_t=0.0025/m$, $a_s=0.075/m$, $N_s=512$ such that $L=38.4/m$.}
 \label{fig:dist_back_latetime_static}
\end{figure}

This simple picture changes rather drastically at late times because of the occurrence of plasma oscillations.
In Fig.~\ref{fig:dist_back_latetime_static} we show the normalized particle number marginal distribution $n^+_\mathcal{P}(p,t)/L$ at different times.
Due to the fact that the electric field changes its sign again and again, the fermions are accelerated back and forth in momentum space over and over again.
The shaking of the fermions by the electric field has several implications:

In contrast to the wedge-shaped spectrum at early times, this results in a peaked $n^+_\mathcal{P}(p,t)/L$ at late times.
It has to be emphasized, however, that this peaked distribution still oscillates around $p=0$ in accordance with the electric field.
Moreover, owing to the ongoing creation of fermion--anti-fermion pairs at times when the electric field reaches its local extrema, the overall magnitude of $n^+_\mathcal{P}(p,t)/L$ increases as well.

\subsection{Space- and time-dependent field}
   
As a further example we consider an inhomogeneous electric background field which is localized in space and time:
\begin{equation}
 \label{fmla_res_efld}
 E(x,t)=E_0\operatorname{sech}^2(\omega t)\exp\left(-\frac{x^2}{2\lambda^2}\right) \ ,
\end{equation}
where $\omega$ and $\lambda$ determine the duration and spatial extent of the pulse, respectively.
Studies based on the continuum Dirac-Heisenberg-Wigner function only recently started to address such inhomogeneous configurations, disregarding the backreaction of created fermion--anti-fermion pairs \cite{Hebenstreit:2011wk,Hebenstreit:2011pm}. Here we are for the first time able to take this fermionic backreaction into account using our lattice techniques. This will allow us to discuss the striking phenomenon of a linear rising potential building up between produced fermion bunches for times exceeding the pulse duration. 

\begin{figure}[b]
 \centering
 \includegraphics[width=0.85\columnwidth]{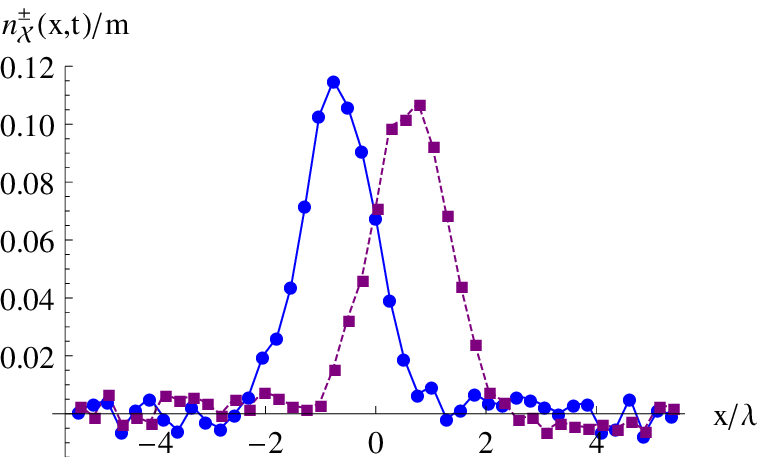}
 \includegraphics[width=0.85\columnwidth]{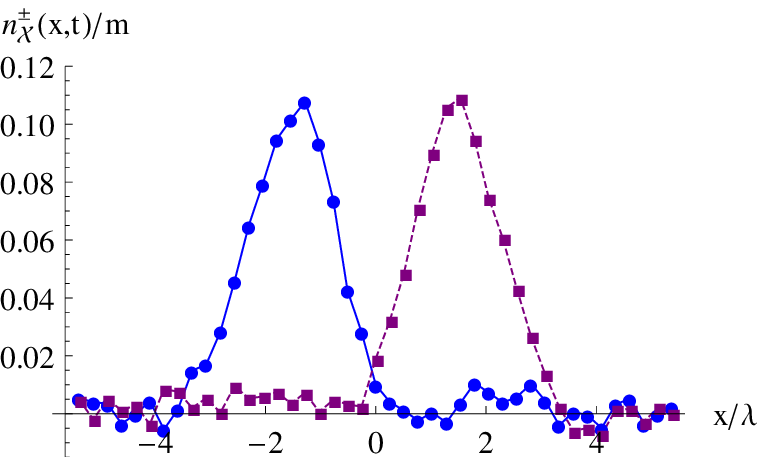}
 \includegraphics[width=0.85\columnwidth]{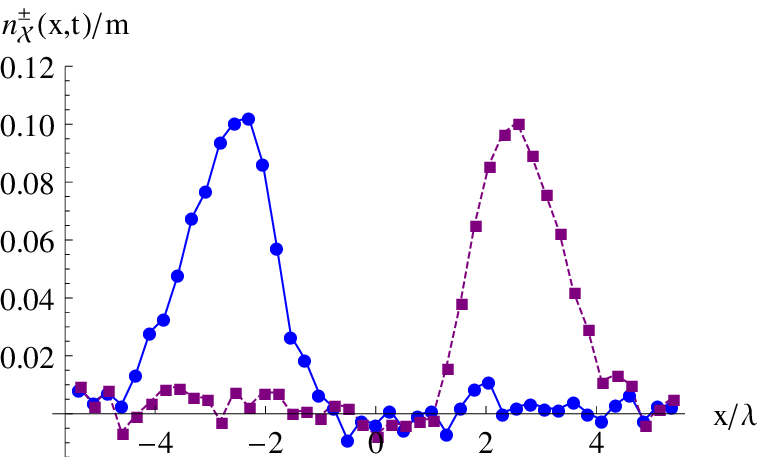}
 \caption{Position space marginal distributions $n^-_\mathcal{X}(x,t)$ (solid line) and $n^+_\mathcal{X}(p,t)$ (dashed line) for $\epsilon=1$ at different times $t=0$ (top), $t=0.6/\omega$ (middle) and $t=1.2/\omega$ (bottom).
 The parameters are $N_\mathrm{sto}=10^5$, $a_t=0.01/m$, $a_s=0.22/m$, $N_s=256$ such that $L=56.32/m$.}
 \label{fig:time_evolution_sech}
\end{figure}

\subsubsection{Particle production without backreaction}

In a first step, we solve the problem without taking into account backreaction. Consequently, we do not evolve the electric field according to (\ref{fmla_th_eom_lat2}) as it does not fulfill Maxwell's equation.
We rather force the electric field to be given according to (\ref{fmla_res_efld}) at every space-time point and investigate the fermion--anti-fermion production in this given background field.
 
In Fig.~\ref{fig:time_evolution_sech} we show the position-space marginal distributions $n^\pm_\mathcal{X}(x,t)$ for three different times, with the electric field parameters $\epsilon=1$, $\omega=0.1m$ and $\lambda=5/m$.
One observes two qualitatively different regimes, corresponding to early times ('creation regime') and late times ('propagation regime').

The fermion--anti-fermion pair creation process takes place at early times, when charged excitations are created in a space region where the electric field acts.
The creation process also comes with a polarization effect, separating positive from negative charges.
It has to be emphasized, however, that $n^+_\mathcal{X}(x,t)$  and $n^-_\mathcal{X}(x,t)$ still overlap at these early times.

This changes in the propagation regime: 
Owing to the acceleration by the electric field, one bunch of excitations with positive charge propagates into the positive x-direction whereas another bunch of excitations with negative charge propagates into the opposite direction. 
Asymptotically, these bunches can be identified with particles and antiparticles, respectively. 

\begin{figure}[b]
 \centering
 \includegraphics[width=\columnwidth]{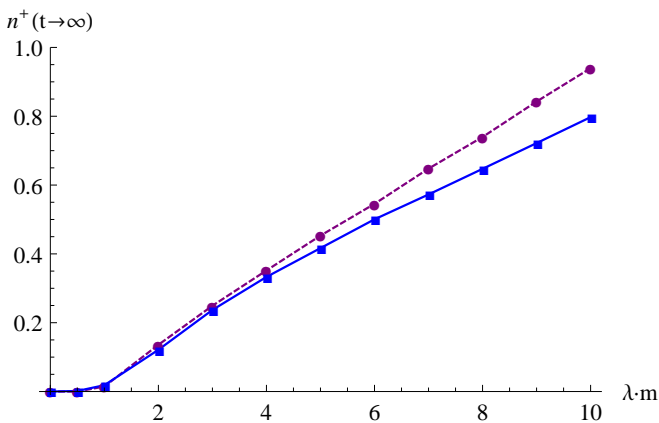}
 \caption{Total number of created particles $n^+(t\to\infty)$ for $\epsilon=1$ as function of the spatial extent $\lambda$ of the pulse. 
 Compared are the full result (solid line) and the result without backreaction (dashed line).
 The parameters are $N_\mathrm{sto}=10^5$, $a_t=0.00125/m$, $a_s=0.22/m$, $N_s=512$ such that $L=112.64/m$.}
 \label{fig:number_sech}
\end{figure}
 
In Fig.~\ref{fig:number_sech} we show the total number of created particles $n^+(t)$ for $t\to\infty$ as a function of the spatial extent $\lambda$. The result without backreaction corresponds to the dashed line. One clearly observes the termination of the fermion--anti-fermion creation process for small values of $\lambda$:
The pair creation process terminates if the work done by the electric field over its spatial extent is too small to provide the rest mass energy of the fermion--anti-fermion pair. This observation is in perfect agreement with previous studies \cite{Dunne:2006ur,Hebenstreit:2011wk,Nikishov:1970br,Gies:2005bz}.
For large values of $\lambda$ we find a linear growth of the particle number, which reflects the scaling of the available electric field energy that grows with $\lambda$.

\subsubsection{Particle production with backreaction}

We now consider the numerical solution of the full lattice problem including backreaction. 
The solid line in Fig.~\ref{fig:number_sech} shows the full result for the total number of created particles $n^+(t\to\infty)$ as a function of the spatial extent $\lambda$. In accordance with the previous discussion we find that the backreaction mechanism tends to decrease the number of created particles since the electric field is weakened by the pair-production. For large enough spatial extent of the pulse, such that the pair-production is significant enough for backreaction to become important, this eventually decreases the slope of the curve growing linearly with $\lambda$ for large spatial extent. 

\begin{figure}[t]
 \centering
 \includegraphics[width=\columnwidth]{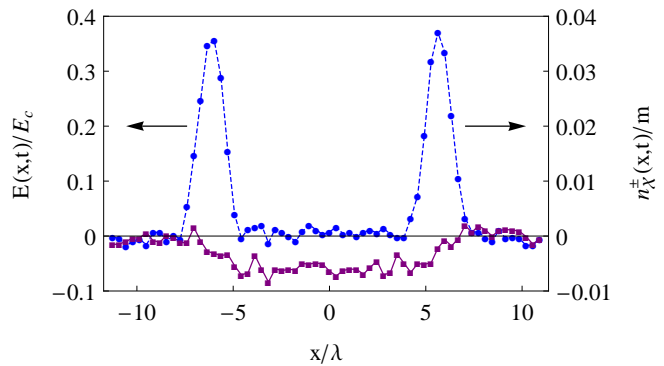}
 \caption{Self-consistent electric field $E(x,t)$ (solid line) and position space marginal distributions $n^\pm_\mathcal{X}(x,t)$ (dashed line) at $t=6/\omega$ for $\epsilon=1$, $\omega=0.2m$ and $\lambda=5/m$.
 The arrows indicate the propagation direction of the particle and anti-particle bunch, respectively.
 The parameters are $N_\mathrm{sto}=10^6$, $a_t=0.0075/m$, $a_s=0.22/m$, $N_s=512$ such that $L=112.64/m$.}
 \label{fig:string_sech}
\end{figure}

The dashed curve in the upper part of Fig.~\ref{fig:string_sech} shows the position-space marginal distribution $n^\pm_\mathcal{X}(x,t)$ at time $t=6/\omega$. 
At this time the duration of the initial electric field pulse and the corresponding pair creation regime is long over. 
The electric field parameters are $\epsilon=1$, $\omega=0.2m$ and  $\lambda=5/m$.

The acceleration by the electric field leads to one bunch of excitations with positive charge propagating into the positive x-direction whereas another bunch of excitations with negative charge is propagating into the opposite direction. 
Most strikingly, we find that a self-consistent electric field $E(x,t)$ between the two fermion bunches builds up in the absence of any external field (\ref{fmla_res_efld}) at these times.
The two bunches consisting of particles and anti-particles act as a capacitor \cite{Chu:2010xc}, 
creating a homogeneous electric field between them whereas there is no field outside them. 
This electric field is shown in the lower part of Fig.~\ref{fig:string_sech}.  
Owing to the description of the fermionic degrees of freedom in terms of low-cost fermions, we observe some small fluctuations in the electric field on top of this homogeneous field. 
Again, these fluctuations decrease with increasing $N_\mathrm{sto}$.

The homogeneous electric field between the fermion bunches can be represented in terms of a linear rising potential.
For larger values of the initial field strength $E_0$ or the coupling $e$, we expect that secondary particle creation due to the self-consistent electric field takes place.
This mechanism would result in the depletion of the electric field reminiscent to the effect of string-breaking. 
This will involve further studies with supercritical initial field strengths which is beyond the scope of the present work and deferred to a future publication.

\section{Conclusion \& outlook}
\label{sec:conclusion}

We investigated fermion--anti-fermion pair production in $1+1$ dimensions based on real-time lattice simulations.
To this end, we discussed the lattice equations of motion using the low-cost fermion algorithm to solve them.
In order to define gauge-invariant fermionic distributions corresponding to charge, energy or particle/anti-particle number, we derived the lattice analogue of the continuum Dirac-Heisenberg-Wigner function. 
In the continuum formulation, gauge invariance of these distributions is achieved by a Wilson line along the straight line. 
On the lattice, however, the straight line path is not unique due the periodicity of the lattice.
We showed that correspondence with established results is achieved by replacing the straight path in the continuum by the shortest path on the lattice. 

Investigating the field-strength dependence of the fermion--anti-fermion production rate in a static background field we accurately reproduced the Schwinger formula.
We then discussed the decay of the field due to the backreaction of the created fermion--anti-fermion pairs.
For the case of inhomogeneous gauge fields we computed for the first time the full problem taking backreaction into account.
Most notably, we could show that the two bunches consisting of particles and anti-particles create a homogeneous electric field between them whereas there is no field outside them.
In subsequent work we will extend these studies to supercritical initial field strengths, which is expected to lead to striking pair creation phenomena reminiscent of string breaking.

It should be emphasized that the real-time lattice simulations are considerably cheaper from a computational point of view than continuum approaches such as based on the Dirac-Heisenberg-Wigner function. 
In view of potential experimental applications it is crucial that strongly inhomogeneous configurations can be well described. 
Strong inhomogeneities are a challenge for alternative approaches based on derivative expansions underlying effective kinetic descriptions. 
Here the lattice approach, which is based on ensemble techniques using inhomogeneous configurations, is particularly powerful.

We employed a low-cost fermion algorithm in our $1+1$ dimensional simulations even though a mode-function expansion of the spinors would have been the more direct way.
One reason for our choice was that we are aiming at investigations of QED in $3+1$ dimensions since then the application of the mode function expansion becomes impracticable.
The anticipated investigations of QED in $3+1$ dimensions will show several major differences compared to the massive Schwinger model.
Most notably, the gauge degrees of freedom are dynamical in contrast to $1+1$ dimensions where the dynamics of the electric field is governed only by the fermionic backreaction.
Moreover, the issue of renormalization will become relevant in $3+1$ dimensions in contrast to the super-renormalizibility of the massive Schwinger model.

By means of the present lattice approach to pair production we hope to deepen our understanding of non-equilibrium QED, including highly topical issues such as threshold lowering, collective phenomena, instabilities or cascades. In the long run, this approach may also be extended to QCD.
Most notably, the investigation of quark--anti-quark production from gluon fields would give important insights into the early stages of relativistic heavy-ion collisions.

\vspace{-0.2cm}
\subsection*{Acknowledgments}

We thank S.~Schlichting and D.~Sexty for helpful discussions as well as R.~Alkofer and H.~Gies for collaborations on related work. 
D.~Gelfand thanks HGS-HIRe for FAIR for support.
F.~Hebenstreit is supported by the Alexander von Humboldt Foundation.


\begin{appendix}

\section{Marginal distributions}
\label{ap:marg}

The definition of the lattice Wigner function (\ref{fmla_th_wig_lat}) is such that we reproduce the continuum expressions for its marginal distributions:
\begin{subequations}
\begin{eqnarray}
 \label{fmla_th_marg_x}
 \mathcal{W}_\mathcal{X}(x,t)&=&-F(x,x;t) \ , \\
 \label{fmla_th_marg_p}
 \mathcal{W}_\mathcal{P}(p,t)&=& -\int{dx_1dx_2e^{-ip(x_1-x_2)}\tilde{F}(x_1,x_2;t)} \ .\  \qquad
\end{eqnarray}
\end{subequations}
               
Regarding the position space marginal distribution on the lattice, we consider:
\begin{equation}
 \mathcal{W}_\mathcal{X}(l,j)=\frac{1}{2L}\sum_{q\in\tilde{\Lambda}_\mathcal{W}}{\mathcal{W}(l,q,j)} \ .
\end{equation}
Upon performing the summation over $q$, we encounter:
\begin{equation}
 \sum_{q\in\tilde{\Lambda}_\mathcal{W}}e^{\pi i (l-2k) q/N_s}=2N_s\delta_{2k,l} \ .
\end{equation}
The Kronecker delta indicates that $\mathcal{W}(l,j)$ is only non-vanishing for even $l$:
\begin{equation}
 \mathcal{W}_\mathcal{X}(l,j)=-F(l,l;j) \ ,
\end{equation}
with $l\in\Lambda$.

Regarding the momentum space marginal distribution on the lattice, we consider:
\begin{equation}
 \mathcal{W}_\mathcal{P}(q,j)=\frac{a_s}{2}\sum_{l\in\Lambda_\mathcal{W}}{\mathcal{W}(l,q,j)} \ .
\end{equation}
Due to the fact that we used the module operation in (\ref{fmla_th_wig_lat}), we obtain:
\begin{alignat}{2}
 \sum_{l=0}^{2N_s-1}&e^{\pi i l q/N_s}\tilde{F}(k,[l-k]_{N_s};j)= \nonumber \\
   &\big(1+e^{i\pi q}\big)\sum_{l=0}^{N_s-1}{e^{\pi i l q/N_s}}\tilde{F}(k,[l-k]_{N_s};j) \ .
\end{alignat}
The factor $(1+e^{i\pi q})$ shows that $\mathcal{W}(q,j)$ is only non-vanishing for even $q$.
Accordingly, if we redefine the summation indices:
\begin{equation}
 l_1=l\in\Lambda \qquad \mathrm{and} \qquad l_2=[l-k]_{N_s}\in\Lambda \ ,
\end{equation}
we reproduce the analogue of (\ref{fmla_th_marg_p}):
\begin{equation}
 \mathcal{W}_\mathcal{P}(q,j)=-a_s^2\sum_{l_1\in\Lambda}\sum_{l_2\in\Lambda}{e^{-2\pi i q (l_1-l_2) / N_s}\tilde{F}(l_1,l_2;j)} \ ,
\end{equation}
with $q\in\tilde{\Lambda}$.

\section{Analytic results for $E(x,t)=E_0$}
\label{ap:stat}

We briefly review some analytic results for the static background field \cite{Hebenstreit:2010vz}.
As a matter of fact, the Dirac equation is analytically solvable for $E(x,t)=E_0$ in terms of parabolic cylinder functions $D_\nu(z)$.
Accordingly, it is also possible to compute $\mathcal{W}(x,p,t)$ explicitly.
\newline
\newline
The pseudo-distributions $m(x,p,t)$, which have been introduced in Sec.~\ref{sec:theory_DHW}, are then given by:
\begin{subequations}
\label{fmla_res_pseud_cont}
\begin{eqnarray}
 \varrho(x,p,t)&=&0 \ , \\
 \epsilon(x,p,t)&=&[2f(p)-1]\omega \ , \\
 n^\pm(x,p,t)&=&f(p) \ .
\end{eqnarray}
\end{subequations}
The function $f(p)$ is usually denotes as the single-particle momentum distribution:
\begin{widetext}
 \begin{equation}
 f(p)=\frac{1}{2}e^{-\pi/4\epsilon}\left[\frac{1}{2\epsilon}\left(1-\frac{p}{\omega}\right)\mathcal{D}_1(p)+\left(1+\frac{p}{\omega}\right)\mathcal{D}_2(p)-\frac{m}{\sqrt{2\epsilon}\,\omega}\mathcal{D}_3(p)\right] \ , 
 \label{fmla_ap_anal}  
 \end{equation}
\end{widetext}
with
\begin{subequations}
\begin{eqnarray}
 \mathcal{D}_1(p)&=&\left|D_{-1+i/2\epsilon}(\hat{p})\right|^2 \ , \\
 \mathcal{D}_2(p)&=&\left|D_{i/2\epsilon}(\hat{p})\right|^2 \ , \\
 \mathcal{D}_3(p)&=&e^{i\pi/4}D_{i/2\epsilon}(\hat{p})D_{-1-i/2\epsilon}(\hat{p}^*) + c.c. \ , \qquad
\end{eqnarray}
\end{subequations}
and
\begin{equation}
 \hat{p}=-\sqrt{\frac{2}{\epsilon}}\frac{p}{m}e^{-i\pi/4} \ .
\end{equation}
We note that $f(p)$ is independent of the time variable $t$. 
It can be shown that $f(p)$ vanishes for small momenta and approaches a non-vanishing constant for large momenta:
\begin{subequations}
\begin{eqnarray}
 \lim_{p\to-\infty}f(p)&=&0 \ , \\
 \lim_{p\to\infty}f(p)&=&\exp\left(-\frac{\pi}{\epsilon}\right) \ .
\end{eqnarray}
\end{subequations}

As the expressions (\ref{fmla_res_pseud_cont}) are spatially homogeneous, they are trivially related to the momentum space marginal distributions $m_\mathcal{P}(p,t)$:
\begin{eqnarray}
 m(x,p,t)=\frac{m_\mathcal{P}(p,t)}{L}  \ , 
\end{eqnarray}
\newline
in the infinite volume $L\to\infty$.
Most notably, the rate at which particles and anti-particles are created is a constant,
so that the total number of particles and anti-particles, respectively, which are created per volume $L$ and time $T$ is given by:
\begin{equation}
 \frac{\Delta n^\pm}{LT}=\frac{eE_0}{2\pi}\exp\left(-\frac{\pi m^2}{eE_0}\right)=\frac{m^2\epsilon}{2\pi}\exp\left(-\frac{\pi}{\epsilon}\right) \ .
\end{equation}

\end{appendix}


\end{document}